\documentclass[sigconf, nonacm=true]{acmart}
\usepackage{kotex}
\usepackage{adjustbox}
\usepackage{subfigure}
\usepackage{framed,multirow}
\usepackage{soul} 

\AtBeginDocument{%
  \providecommand\BibTeX{{%
    \normalfont B\kern-0.5em{\scshape i\kern-0.25em b}\kern-0.8em\TeX}}}

\newcommand\HH{
  \global\let\savedtextbullet\textbullet
  \gdef\textbullet{%
    \par\noindent\savedtextbullet\global\let\textbullet\savedtextbullet
  }%
}

\copyrightyear{2022}
\acmYear{2022}
\setcopyright{acmcopyright}\acmConference[SIGIR '22]{Proceedings of the 45th International ACM SIGIR Conference on Research and Development in Information Retrieval}{July 11--15, 2022}{Madirid, Spain}
\acmBooktitle{Proceedings of the 45th International ACM SIGIR Conference on Research and Development in Information Retrieval (SIGIR '22), July 11--15, 2022, Madrid, Spain}
\acmPrice{15.00}
\acmDOI{}
\acmISBN{}

\begin{document}
\fancyhead{}

\title{Finding Inverse Document Frequency Information in BERT}

\author{Jaekeol Choi}
\authornote{Both authors contributed equally to this research.}
\affiliation{%
	\institution{Seoul National University \\ \& Naver Corp.}
}
\email{jaekeol.choi@snu.ac.kr}

\author{Euna Jung}
\authornotemark[1]
\affiliation{%
	\institution{GSCST \\ 
	Seoul National University}
	\streetaddress{}
}
\email{xlpczv@snu.ac.kr}

\author{Sungjun Lim}
\affiliation{%
	\institution{Chung-Ang University}
	\streetaddress{}
}
\email{lsjung567@naver.com}

\author{Wonjong Rhee}
\affiliation{%
	\institution{GSCST, GSAI, AIIS \\
	Seoul National University}
}
\email{wrhee@snu.ac.kr}


\begin{abstract}
For many decades, BM25 and its variants have been the dominant document retrieval approach, where their two underlying features are Term Frequency~(TF) and Inverse Document Frequency~(IDF). The traditional approach, however, is being rapidly replaced by Neural Ranking Models~(NRMs) that can exploit semantic features. In this work, we consider BERT-based NRMs and study if IDF information is present in the NRMs. This simple question is interesting because IDF has been indispensable for the traditional lexical matching, but global features like IDF are not explicitly learned by neural language models including BERT. We adopt linear probing as the main analysis tool because typical BERT based NRMs utilize linear or inner-product based score aggregators. We analyze input embeddings, representations of all BERT layers, and the self-attention weights of CLS. By studying MS-MARCO dataset with three BERT-based models, we show that all of them contain information that is strongly dependent on IDF.\footnote{The code will be available after the review}

\end{abstract}





\maketitle
\section{Introduction}
\label{sec:intro}
Since its inception in 1970's, BM25~\cite{robertson2009probabilistic} has been one of the most popular ranking functions for search engines. For a given search query $\mathrm{q}$, BM25 evaluates the relevance score of a candidate document $\mathrm{d} \in \mathrm{D}$ as
\begin{equation}
    \label{eq:idf_tf}
    \mathsf{score}(\mathrm{q},\mathrm{d}) = \sum_{i=1}^{N}{\mathsf{IDF}(q_{i},\mathrm{D}) \cdot \mathsf{TF}(q_{i},\mathrm{d})},
\end{equation}
where $\mathrm{D}$ is the set of candidate documents, $\mathrm{q}$ contains $N$ query terms $\{q_{1}, q_{2}, \cdots, q_{N}\}$,
$\mathsf{TF}(q_{i}, \mathrm{d})$ is the \textit{Term Frequency}~(TF) that is the frequency of $q_{i}$ within the candidate document $\mathrm{d}$, and 
$\mathsf{IDF}(q_{i},\mathrm{D})$ is the \textit{Inverse Document Frequency}~(IDF) that is the inverse of the term $q_{i}$'s frequency over the document set $\mathrm{D}$.
Both TF and IDF have been successfully utilized for lexical matching algorithms such as BM25. The two, however, have quite different characteristics because TF is a local feature and IDF is a global feature. IDF reflects global information over the entire document set $\mathrm{D}$ while TF is a simple similarity measure between the pair $\mathrm{q}$ and $\mathrm{d}$.

Since the advent of BERT~(Bidirectional Encoder Representations from Transformers~\cite{devlin2019bert}), a variety of BERT-based Neural Ranking Models~(NRMs) have been proposed~\cite{nogueira2019passage, khattab2020colbert}. A BERT-based NRM relies on the general-purpose language representation capability of BERT. For the specific goal of document ranking, a score aggregator\footnote{Here, by an aggregator we refer to a model that estimates a relevance score using BERT representations.} is cascaded where it typically performs only a simple operation such as a linear regression or an inner-product. Despite the simplicity, NRMs can easily outperform the traditional lexical matching models. As in many other natural language tasks, the outstanding performance is known to be the result of BERT's excellent representations. On the other hand, BERT is trained only with sentence-level tasks such as Masked Language Modeling~(MLM). This means a global feature like IDF cannot be learned explicitly because its calculation requires an inspection of the entire document set $\mathrm{D}$. Considering that IDF has been a pivotal feature for document ranking, it becomes natural to ask if IDF information is available in NRMs and if we can improve NRMs using global features like IDF. 

In this work, we focus on IDF where the ground-truth IDF values of the tokens are pre-calculated using $\mathrm{D}$. The ground-truth values are used only for the investigations and they do not play any part in BERT-based NRMs. We first investigate if the ground-truth IDF values 
can be extracted from the input embeddings or from the BERT layer representations. A basic and popular probing technique known as linear probing~\cite{alain2016understanding} is adopted to avoid ambiguous interpretations -- if IDF information can be reliably extracted with a linear probing, it indicates that IDF information is not only present but also linearly decodable from the embeddings or representations. 
Additionally, we investigate the correlation between $\mathrm{q}$'s IDF vector and the self-attention weights that are used to form CLS in each head. If a high correlation is found for a head, it 
indicates that a strong linear dependency exists between $\mathsf{IDF}(q_{i},\mathrm{D})$ and token $i$'s attention value. The CLS vector of such a head can be interpreted as an IDF-weighted sum of the token representations.

\begin{figure*}[t]
    \centering{
    \hfill
    \subfigure[Evaluation of IDF probing]
    {\includegraphics[width=0.43\linewidth]{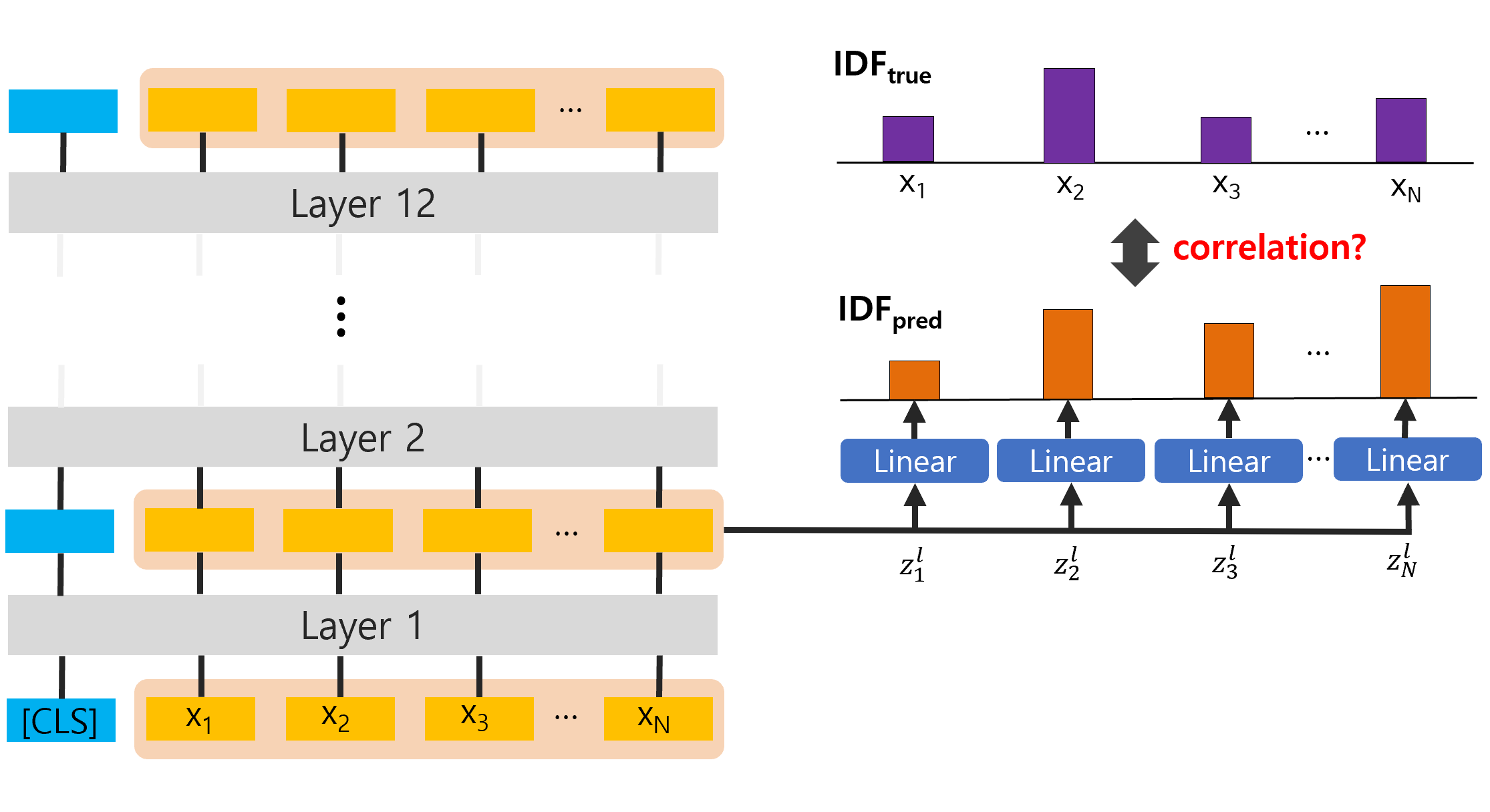}}
    \hfill
    \subfigure[CLS attention analysis] {\includegraphics[width=0.43\linewidth]{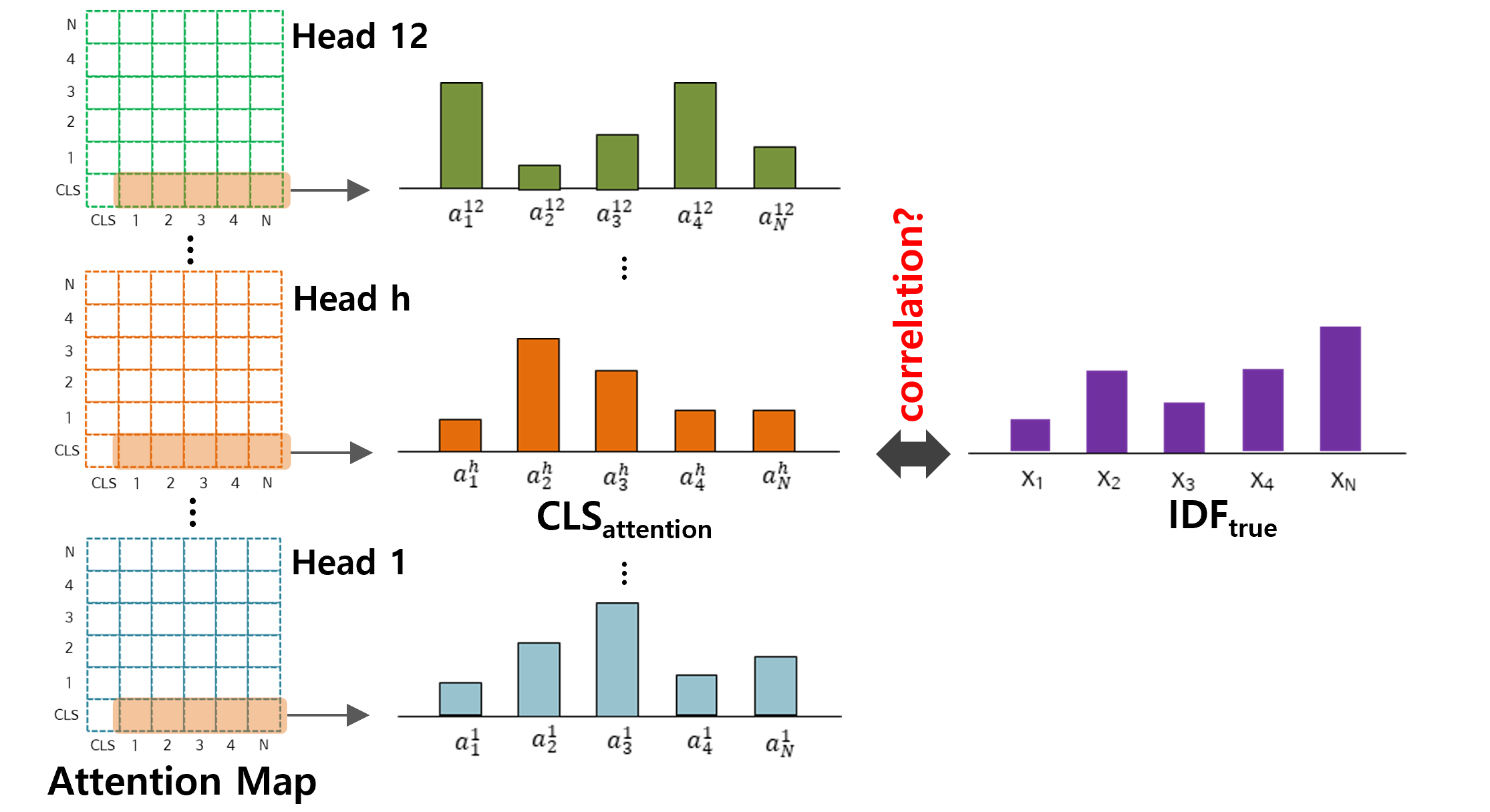}}
    \hfill 
    }
    \caption{Analysis methods: (a) For IDF probing, a linear model is trained to predict the ground-truth IDF value of each token. For test data, Pearson correlation is evaluated between the prediction vector and the ground-truth vector. This is repeated over all BERT layers. 
    (b) For CLS attention analysis, the attention weight vector of each CLS head is directly analyzed where Pearson correlation is evaluated between the attention weight vector and the IDF's ground-truth vector.}
    \label{fig:method}
\end{figure*}

We performed our experiments with MS-MARCO, a large-scale document ranking dataset. For the NRM's language model, we considered pre-trained BERT, fine-tuned ColBERT~\cite{khattab2020colbert}, and prefix-tuned ColBERT~\cite{jung2021semi}. ColBERT is known as a state-of-the-art bi-encoder NRM, and prefix-tuned ColBERT is known to outperform ColBERT. Through experiments, we show that IDF information can be reliably extracted from BERT-based NRMs. This does not necessarily mean an NRM explicitly calculates or utilizes IDF, but it confirms that some information within NRM is strongly correlated with IDF. Furthermore, we found that IDF information can be more reliably extracted from BERT's input embeddings than from Word2Vec~\cite{mikolov2013efficient} or Glove~\cite{pennington2014glove} embeddings, and from the representations of a better performing NRM model than from a worse performing one. Overall, availability of IDF information correlated well with the performance of information retrieval. As for the attention weight vectors to form CLS, we found that the attention vectors of some heads strongly correlate with the IDF vector. BERT heads are known to perform different roles and our result indicates that some of the heads might play more IDF-like roles than the other heads.

\section{Methodology}
To examine if IDF information is available in BERT, we employ IDF probing and CLS attention weight correlation for the analysis. 


\subsection{IDF Probing}
The ground-truth IDF values are pre-computed using the document set of MS-MARCO. For a given token $\mathbf{x}$ and the document set $\mathsf{D}$, the true IDF is defined as below.
\begin{equation}
     \mathrm{IDF_{true}}(\mathbf{x}, \mathsf{D}) =  -\log P(\mathbf{x}|\mathsf{D})
\end{equation}
For an input query with $N$ tokens, we will note its $i$'th token as $\mathbf{x}_i$. Note that $\mathbf{x}_i$ is equivalent to the query term $q_i$ in Section~\ref{sec:intro}.
%
For the layer $l$ of BERT, the IDF value of $i$'th representation token $\mathbf{z}_i^l$ is predicted with a linear model. The linear probing is modeled as
\begin{equation}
    \mathrm{IDF}^{l}_\mathrm{pred}(\mathbf{z}_i^l) = {\mathbf{w}}^l \cdot \mathbf{z}_i^l + b^l,
\end{equation}
where $\mathbf{w}^l$ and $b^l$ are the linear weight parameters and the bias parameter of the linear model, respectively. 
The query set of MS-MARCO is divided into training and test datasets, and the linear model is learned with the training dataset using the MSE loss for each query as
\begin{equation}
    \mathcal{L} = \frac{1}{N} \sum_{i=1}^N   \left(\mathrm{IDF_{true}}(\mathbf{x}_i, \mathsf{D}) -  \mathrm{IDF}_{\mathrm{pred}}^l(\mathbf{z}_i^l) \right)^2,
\end{equation}
where $\mathrm{IDF_{true}}(\mathbf{x}_i, \mathsf{D})$ is the ground-truth IDF value of the token $\mathbf{x}_i$. 
After the training, we evaluate each BERT layer's IDF prediction performance using the test dataset where the Pearson correlation~\cite{benesty2009pearson} is calculated between the ground-truth IDF vector of size $N$ with its $i$'th element $\mathrm{IDF_{true}}(\mathbf{x}_i, \mathsf{D})$ and the predicted IDF vector of size $N$ with its $i$'th element $\mathrm{IDF}_{\mathrm{pred}}^l(\mathbf{z}_i^l)$.
The IDF probing is shown in Figure~\ref{fig:method}(a).

\subsection{CLS Attention Weight Analysis}
[CLS] is a special token for aggregating information of all token representations, and the last layer's [CLS] is commonly used for sentence-level classification tasks. 
By analyzing the self-attention weights of the last BERT layer, we can inspect how each head's CLS representation is related to the ground-truth IDF vector. For the head $h$, the last layer's [CLS] is defined using the self-attention weight vector with its $i$'th element $a^h_i$. In our study, the Pearson correlation of the head $h$ is calculated between the ground-truth IDF vector with its $i$'th element $\mathrm{IDF_{true}}(\mathbf{x}_i, \mathsf{D})$ and the head $h$'s CLS attention weight vector with its $i$'th element $a^h_i$. The CLS attention weight analysis is shown in Figure~\ref{fig:method}(b). We evaluate each head as the average correlation over all the queries in the test dataset.

\section{Experimental Result}

\subsection{Experimental Setup}
\subsubsection{Dataset and Metric}
We conduct our experiments on MS-MARCO dataset\footnote{3.2M documents, 372K queries, https://microsoft.github.io/msmarco/TREC-Deep-Learning-2019} for document ranking.
We pre-compute the ground-truth IDF value for each token using the entire document set $\mathrm{D}$, and we use queries of MS-MARCO as the input sequences. We split the tokens in the query dataset into training, validation, and test. Only the test dataset is used for the evaluations. For a consistent and intuitive analysis, we evaluate Pearson correlation against the ground-truth IDF for both IDF probing and CLS attention analysis.

\subsubsection{Embeddings and BERT models}
We first explore IDF probing performance of random, Word2Vec~\cite{mikolov2013efficient}, GloVe~\cite{pennington2014glove}, and three BERT embeddings. For a fair comparison, we trained Word2Vec and GloVe using WordPiece\footnote{Vocab size=30,522} vocabulary in the dimension of 768 as in BERT. Then, we examine BERT representations of three different BERT-based NRM models. All three are based on an original BERT model without any weight modification and it corresponds to the Pre-trained model. Fine-tuned model and Prefix-tuned model are further fine-tuned for document ranking with ColBERT~\cite{khattab2020colbert} architecture and MS-MARCO dataset. The Prefix-tuned model is based on the prefix-tuning method described in~\cite{li2021prefix, jung2021semi}.

\subsubsection{Training and optimization}
Our experiments are implemented in Python3 and Pytorch1, and we use the transformer library provided by Hugging Face. To train our linear model of IDF probing, we used Adam optimizer with learning rate of 0.00005, batch size of 128, and maximum epoch of 100. When fine-tuning the Fine-tuned and Prefix-tuned models, we followed \cite{jung2021semi}.

\begin{table}[t]
    \centering
    \caption{IDF probing over six types of embeddings. Each embedding has a dimension of 768.}
    \adjustbox{max width=\linewidth}{%
    \begin{tabular}{c|c|c|c}
        \toprule
        Embedding & Correlation & Embedding & Correlation\\
        \midrule
        Random & 0.036~($\pm$0.223) & BERT~(Pre-trained) & 0.970~($\pm$0.005)  \\
        Word2Vec\cite{mikolov2013efficient} & 0.783~($\pm$0.092) & BERT~(Fine-tuned) & 0.971~($\pm$0.005) \\
        GloVe\cite{pennington2014glove} & 0.948~($\pm$0.012) & BERT~(Prefix-tuned) & 0.970~($\pm$0.005) \\
        \bottomrule
    \end{tabular}}
    \label{tab:result_emb_probing}
\end{table}

\begin{table}[t]
	\caption{IDF probing over all BERT layers. Three BERT-based NRM models are investigated.
	}
	\label{tab:result_probing}
	\adjustbox{max width=\linewidth}{%
	\begin{tabular}{c|rrr|c|rrr}
		\toprule
		\multirow{2}{*}{Layer} & \multicolumn{1}{c}{Pre-} & \multicolumn{1}{c}{Fine-} & \multicolumn{1}{c|}{Prefix-} & \multirow{2}{*}{Layer} & \multicolumn{1}{c}{Pre-} & \multicolumn{1}{c}{Fine-} & \multicolumn{1}{c}{Prefix-} \\
		& \multicolumn{1}{c}{trained} & \multicolumn{1}{c}{tuned} & \multicolumn{1}{c|}{tuned} & & \multicolumn{1}{c}{trained} & \multicolumn{1}{c}{tuned} & \multicolumn{1}{c}{tuned} \\
		\midrule
		Layer 1 & 0.962 & 0.962 & 0.965 & Layer 7 &0.914 & 0.930& 0.939\\
		Layer 2 & 0.959 & 0.961 & 0.963 & Layer 8 & 0.900 & 0.920 & 0.934\\
		Layer 3 & 0.955 & 0.958 & 0.959 & Layer 9 & 0.884 & 0.905 & 0.927 \\
		Layer 4 & 0.945 & 0.953 & 0.954 & Layer 10 & 0.868 & 0.889 & 0.918 \\
		Layer 5 & 0.936 & 0.946 & 0.948 & Layer 11 & 0.848 & 0.868 & 0.910\\
		Layer 6 & 0.927 & 0.940 & 0.946 & Layer 12 & 0.802 & 0.823 & 0.895 \\
       \bottomrule
	\end{tabular}}
\end{table}

\begin{figure}[t]
    \centering
    \subfigure[Embeddings] {\includegraphics[width=0.47\linewidth]{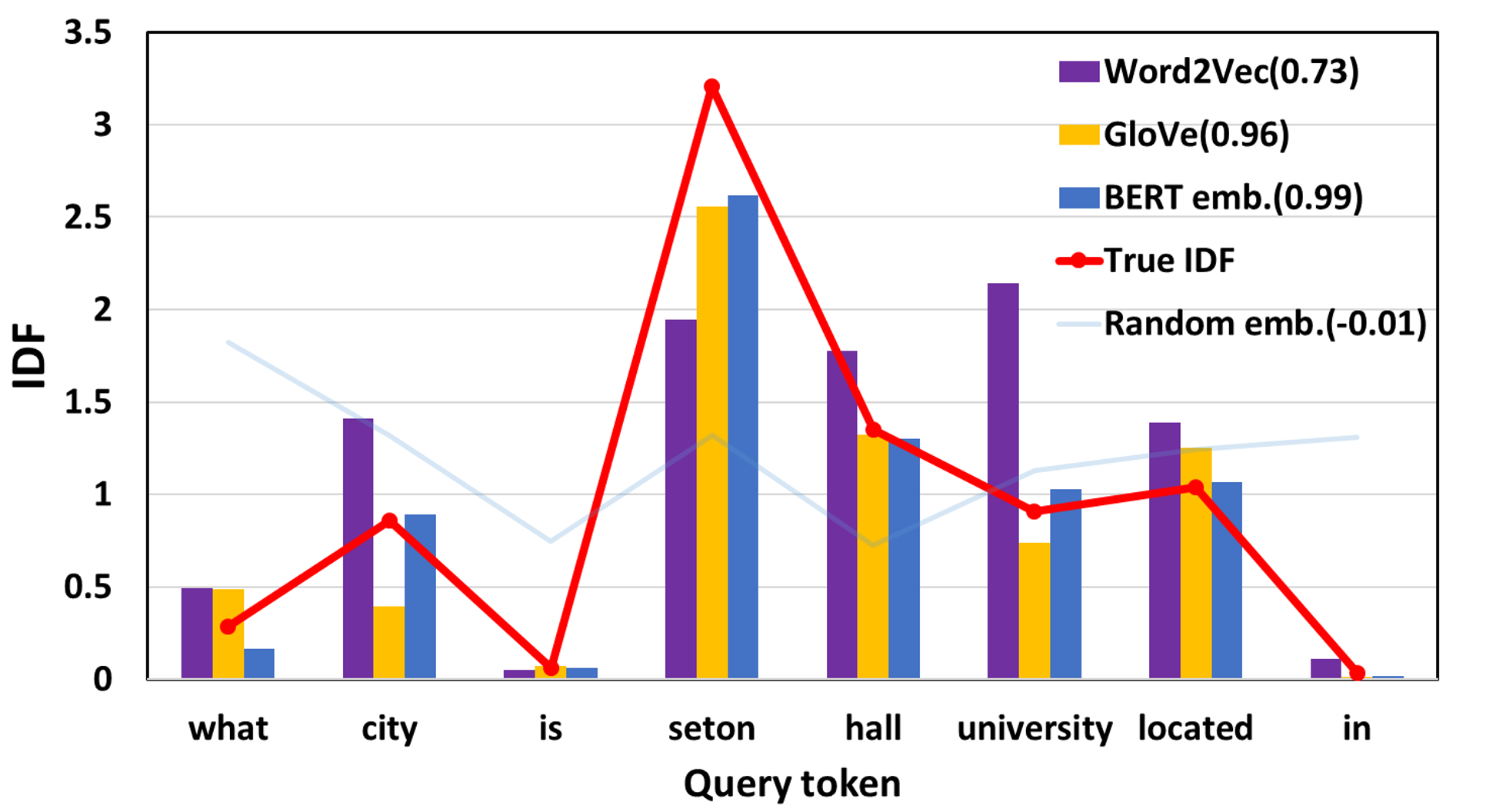}}
    \subfigure[Layer12 representations]{\includegraphics[width=0.47\linewidth]{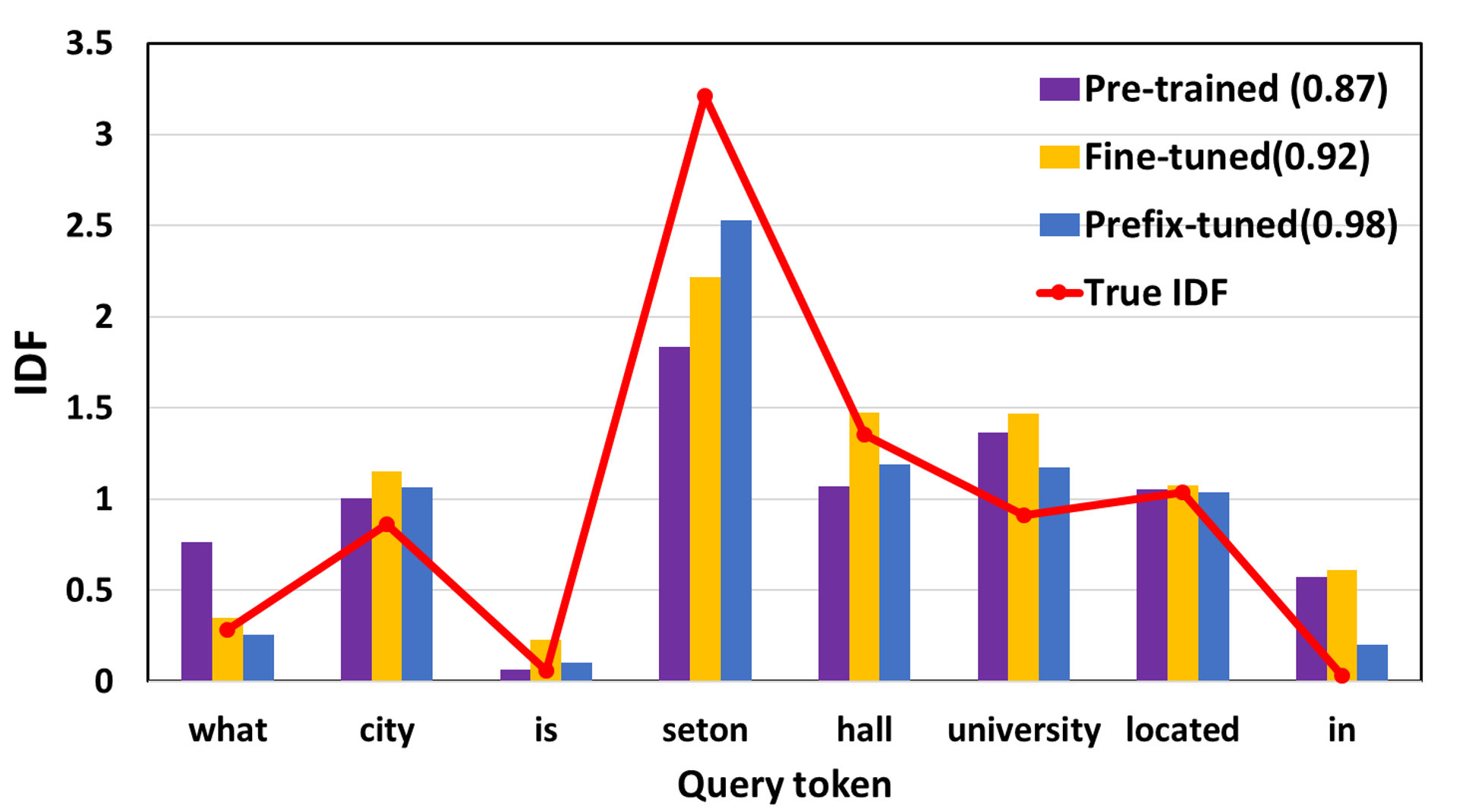}}
    \caption{An example case of IDF probing~(QID:596146).  
    (a) The predicted IDF values are shown for four types of embeddings. The true IDF values are shown together.
    (b) The predicted IDF values from the Layer12 representations for three different models.}
    \label{fig:probing_example}
\end{figure}

\subsection{IDF Probing Result}
We analyze IDF information in embeddings, layers, and models.

\subsubsection{Embeddings}
IDF probing performance for six types of embeddings are shown in Table \ref{tab:result_emb_probing}. As expected, randomly generated embeddings make it impossible to predict the ground-truth IDF, and the resulting correlation to the ground-truth IDF vector is close to zero. Word2Vec learns its embeddings without any explicit procedure for learning global features, but its probing results show a high correlation with the ground-truth IDF at 0.783. GloVe results in 0.948 that is significantly larger than Word2Vec, and it is probably because GloVe utilizes the probability of the entire corpus during its learning. As in Word2Vec, BERT embeddings are learned without any explicit procedure for learning global features. Nonetheless, all three BERT models result in very high correlations around 0.970. An example case is shown in Figure~\ref{fig:probing_example}(a). 
The example case was randomly sampled from the queries with 10\% or more difference in correlation between the Pre-trained BERT and the Prefix-tuned model. 

\subsubsection{BERT layers}
IDF probing performance over the twelve BERT layers is shown in Table \ref{tab:result_probing}. For the Pre-trained BERT, it can be observed that the correlation monotonically decreases as the layer number increases. The correlation drops by 17\% from the first layer to the last layer. The monotonic decrease can be also observed from the Fine-tuned model and the Prefix-tuned model. The behavior indicates that linear probing performance is not improved by BERT's local operations as in the data processing inequality of information theory~\cite{cover1999elements}. Therefore, we can infer that IDF information is not only learned and stored in BERT's input embeddings but also stored in a linearly decodable way. By locally processing a query's input embeddings toward the upper layers, we can only lose IDF information that is linearly useful for information retrieval tasks.

\subsubsection{BERT models}
In Table \ref{tab:result_probing}, IDF probing performance over the three BERT-based NRMs can be found in addition to the layer-wise evaluations. While the monotonic decrease over the layers is a common behavior for the three models, the severity is different. The drop is 17\% for Pre-trained BERT, but it is 14\% for Fine-tuned BERT and only 7\% for Prefixed-tuned BERT. It is interesting to note that the BERT weights of Fine-tuned model and Prefixed-tuned model were updated using MS-MARCO's document ranking task. The post-training has clearly affected the BERT weights in a way that makes IDF information retained better in the upper layers.

\subsection{CLS Attention Analysis Result}

\begin{figure}[t]
    \centering
    \subfigure[Pre-trained BERT]{\includegraphics[width=0.47\linewidth]{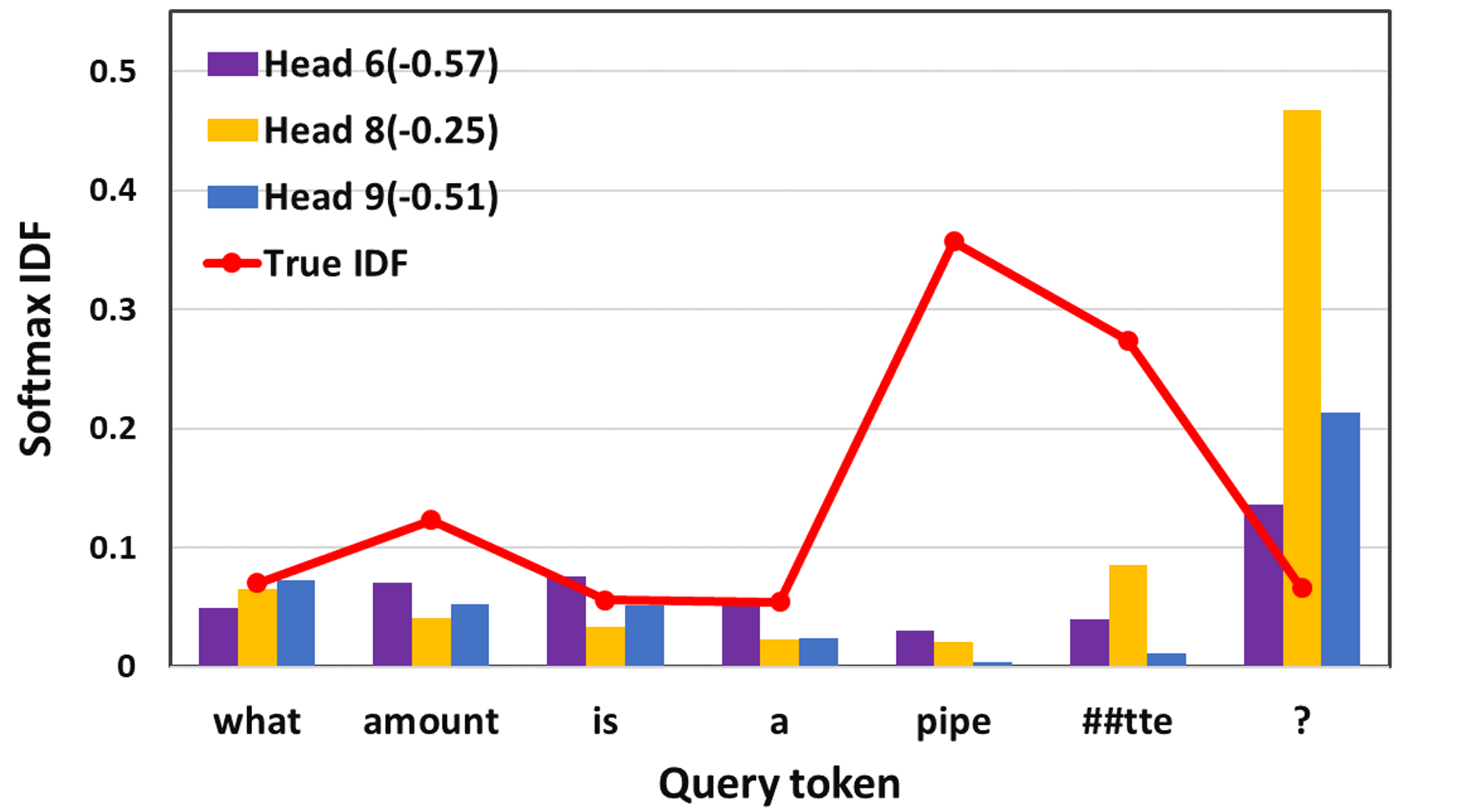}}
    \subfigure[Fine-tuned BERT]{\includegraphics[width=0.47\linewidth]{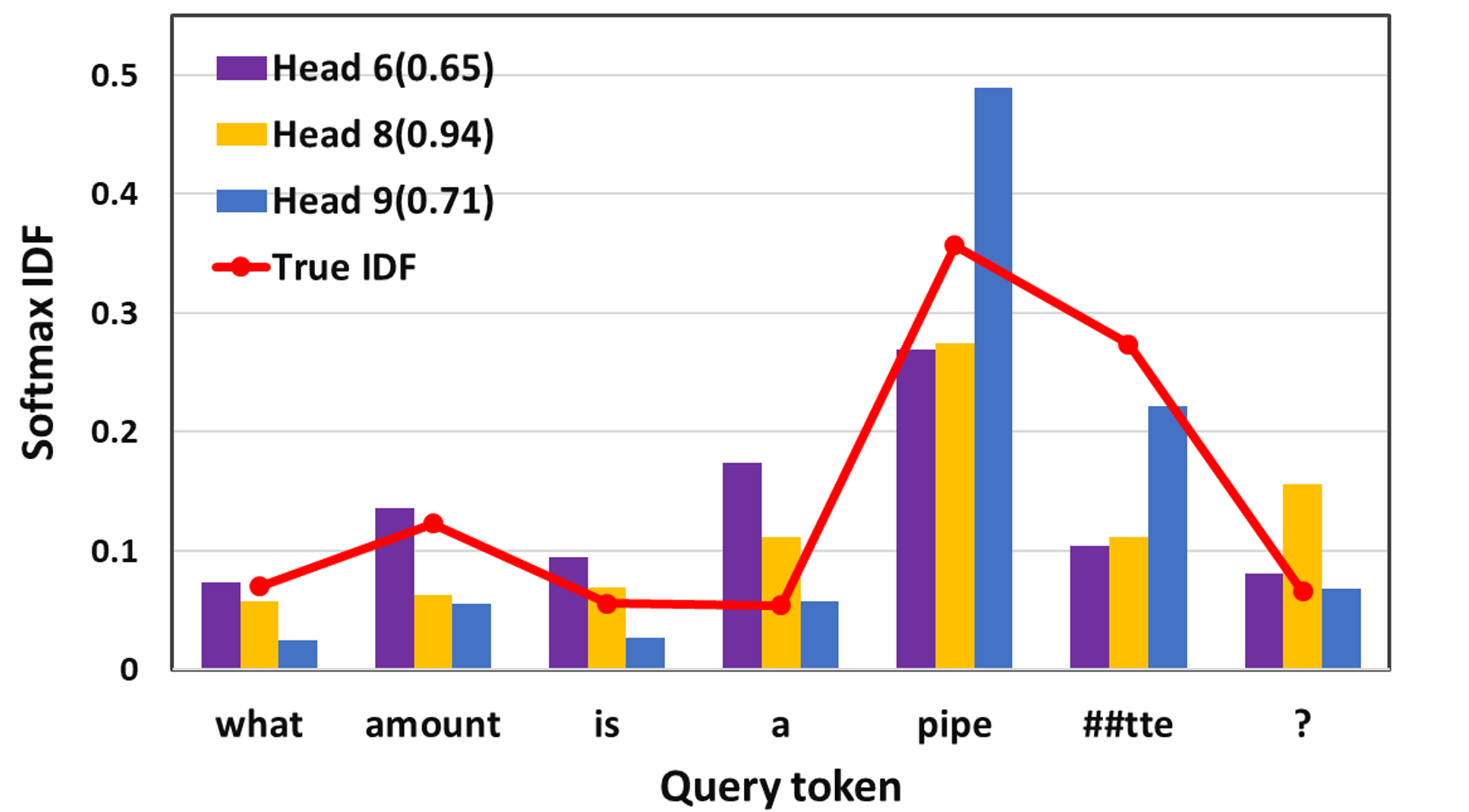}}
    \newline
    \subfigure[Pre-trained BERT ]{\includegraphics[width=0.47\linewidth]{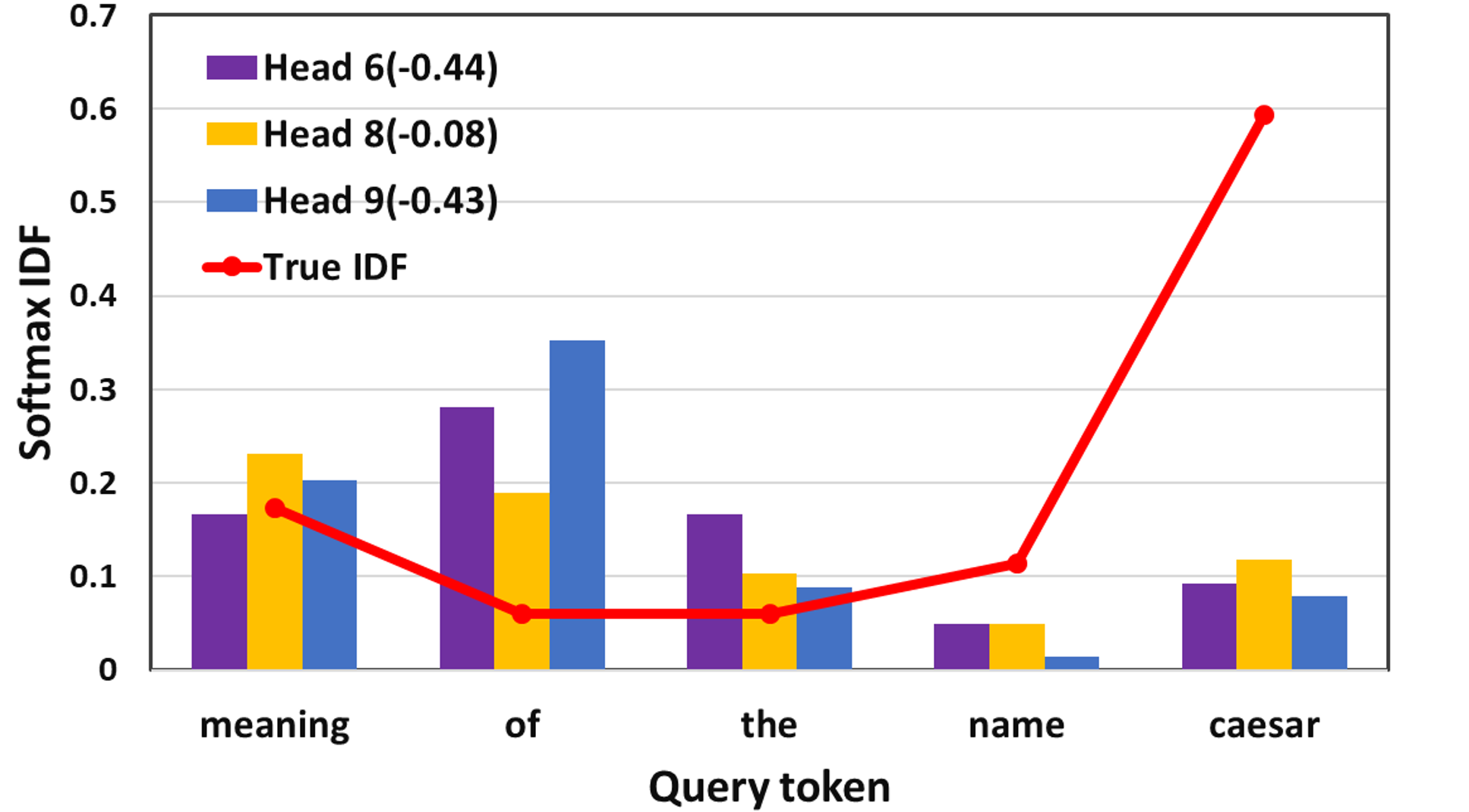}}
    \subfigure[Prefix-tuned BERT]{\includegraphics[width=0.47\linewidth]{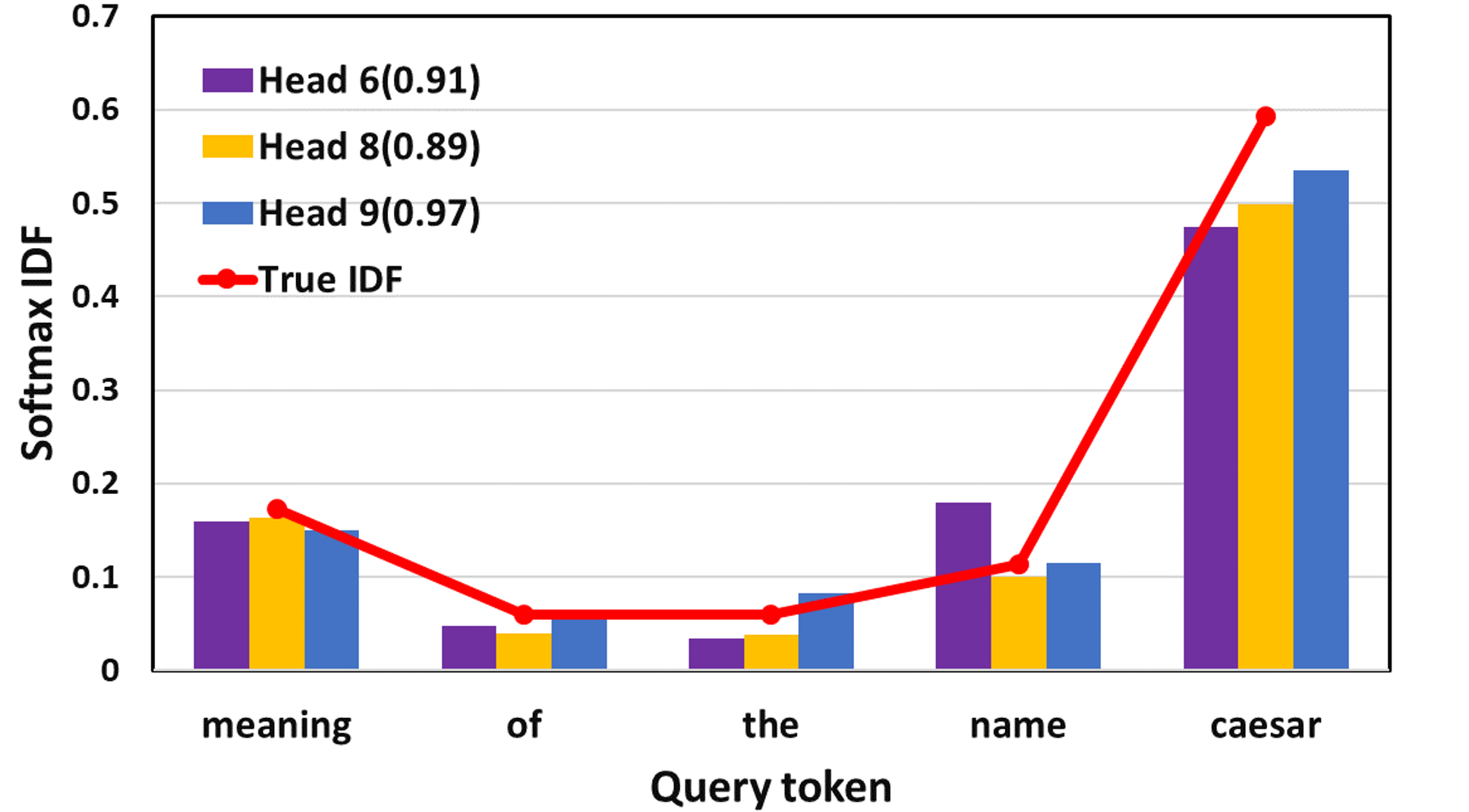}}
    \caption{Two example cases of CLS attention weight analysis. 
    For the heads with correlation above 0.3, their attention weights are shown.
    (a) and (b): Pre-trained BERT vs. Fine-tuned BERT (QID:449647).
    (c) and (d): Pre-trained BERT vs. Prefix-tuned BERT (QID:1167052).
    }
    \label{fig:cls_atten_example}
\end{figure}

Table \ref{tab:result_cls} shows the correlation between the last BERT layer's CLS attention weights and the ground-truth IDF vector. The average correlation values were evaluated over the test dataset of queries.
For the Pre-trained BERT, most of the heads correlate negatively with the ground-truth IDF. A negative correlation indicates that a CLS attention weight tends to be large for a frequently used token and small for an infrequently used token. Even the largest positive correlation is only 0.078 (Head 8) that is close to zero. 
On the other hand, Fine-tuned and Prefixed-tuned models behave differently. Some of the heads show large positive correlations, up to 0.531, indicating that the CLS representation is a weighted sum of the token representations where the weight vector is similar to the ground-truth IDF vector. Therefore, the corresponding heads (heads 6, 8, and 9) can be related to the score function in Equation~(\ref{eq:idf_tf}). Looking at the negatively correlated heads, it can be observed that most of them show even more negative correlations than in the Pre-trained BERT. This indicates that such heads might play different roles than the positively correlated heads -- roles that cannot be implemented in TF-IDF scoring. 
Two example cases are shown in Figure \ref{fig:cls_atten_example}. Clearly, Pre-trained BERT's behavior is in contrast to the Fine-tuned or Prefixed-tuned BERT's behavior. Note that we have performed a light screening to choose the two cases. 

 \begin{table}[t]
	\caption{Correlation between the ground-truth IDF vector and the last layer's CLS attention weight vector.}
	\label{tab:result_cls}
	\adjustbox{max width=\linewidth}{%
	\begin{tabular}{c|rrr|c|rrr}
		\toprule
		\multirow{2}{*}{Head} & \multicolumn{1}{c}{Pre-} & \multicolumn{1}{c}{Fine-} & \multicolumn{1}{c|}{Prefix-} & \multirow{2}{*}{Head} & \multicolumn{1}{c}{Pre-} & \multicolumn{1}{c}{Fine-} & \multicolumn{1}{c}{Prefix-} \\
		& \multicolumn{1}{c}{trained} & \multicolumn{1}{c}{tuned} & \multicolumn{1}{c|}{tuned} & & \multicolumn{1}{c}{trained} & \multicolumn{1}{c}{tuned} & \multicolumn{1}{c}{tuned} \\
		\midrule
		Head 1 & -0.209 & -0.336 & 0.114 & Head 7 & -0.343 & -0.550 & -0.445  \\
		Head 2 & -0.137 & -0.547 & -0.470 & \bf{Head 8} & 0.078 & \bf{0.440} & \bf{0.531}  \\
		Head 3 & -0.322 & -0.590 & -0.516 & \bf{Head 9} & -0.278 & \bf{0.513} & \bf{0.501}  \\
		Head 4 & -0.318 & -0.506 & -0.498 & Head 10 & -0.351 & -0.469 & -0.524 \\
		Head 5 & -0.326 & -0.546 & -0.430 & Head 11 & -0.252 & -0.552 & -0.350 \\
		\bf{Head 6} & 0.059 & 0.126 & \bf{0.340} & Head 12 & -0.051  &  0.136 & -0.154  \\
       \bottomrule
	\end{tabular}}
\end{table}

\section{Discussion}

\textbf{Finding IDF information in BERT:}
Based on our empirical investigations, we can conclude that IDF information can be reliably found in BERT representations, especially in the embeddings. This is interesting because BERT neither explicitly learns global features at the time of training nor uses any global operation at the time of inference. 
When BERT is fine-tuned for document ranking, we have found that the weights of BERT are tuned in a way that reduces the loss of IDF information in the upper layers.  
Furthermore, we have found that CLS attention weights of some heads become more relevant to IDF by the fine-tuning. 
It is noted that there have been numerous studies on the roles of heads. For instance, \citet{voita2019analyzing} discovered a special head that points to the least frequent tokens.

\textbf{Exploiting IDF information for performance improvement:}
For any document ranking dataset, the ground-truth IDF values can be easily calculated. Therefore, we can consider using the true IDF values for improving the performance of BERT-based NRMs. We have tried a few simple methods where IDF is used to regularize the fine-tuning. For instance, we have tried a penalty regularization that aims to improve the linear probing performance of the upper layers. All of our simple methods, however, failed to improve the document ranking performance. The result might be due to the limitations of the methods that we have tried, but there might be a chance that such methods are not helpful because a pre-trained BERT already has a significant amount of IDF information.

\textbf{Utilizing lower-layer representations:}
BERT performs local operations at the time of inference and we have shown that the amount of linearly decodable IDF information decreases with the increase in layer number. A simple work-around for this problem is to use the representations of lower layers as well as the representations of the last layer for the relevance score calculation. In this case, it can be hypothesized that lower layers provide lexical and IDF-like information while the upper layers provide semantic and syntactic information \cite{rogers2020primer}. CEDR~\cite{macavaney2019cedr} reported that a ranking model using BERT representations of all 12 layers can outperform the basic model that utilizes only the last layer's representations.

\section{Conclusion}
In this study, we have investigated if IDF information is present in BERT and BERT-based NRMs. 
Using a probing technique, we have shown that IDF information can be extracted from BERT representations especially in the lower layers. Fine-tuning for document ranking has an effect of increasing the amount of IDF information in the upper layers. Fine-tuning also made CLS attention weights of three heads positively dependent to the IDF information. 
Our results show that we can certainly find IDF information in BERT, especially when BERT is fine-tuned for document ranking.

\bibliographystyle{ACM-Reference-Format}
\bibliography{acmart}


\end{document}